\documentclass[aps,pra,twocolumn,superscriptaddress,notitlepage,nofootinbib,longbibliography]{revtex4-2}
\usepackage{mathrsfs}
\usepackage{epsfig}
\usepackage{graphicx}
\usepackage{amsfonts}
\usepackage{dcolumn}
\usepackage[figuresright]{rotating}
\usepackage{amssymb}
\usepackage{amsmath}
\usepackage{dcolumn}
\usepackage[dvipsnames]{xcolor}
\usepackage{bm}
\usepackage{xcolor}
\usepackage{color}
\usepackage{braket}
\usepackage{units}
\usepackage{xspace}

\usepackage{bbold}
\definecolor{mypink3}{cmyk}{0, 0.7808, 0.4429, 0.1412}
\definecolor{mypink1}{rgb}{0.858, 0.188, 0.478}
\definecolor{mypink2}{RGB}{219, 48, 122}
\usepackage[colorlinks, linkcolor=blue, citecolor=blue, urlcolor=blue, breaklinks=true]{hyperref}
\definecolor{lapislazuli}{rgb}{0, 0, 1}
\definecolor{YKblue}{rgb}{0.0, 0.18, 0.65}
\definecolor{carmine}{rgb}{0.81, 0.09, 0.03}
\definecolor{lavender}{rgb}{0.84, 0.49, 0.87}
\usepackage{nicefrac,xfrac}
\usepackage[normalem]{ulem}

\usepackage[fleqn]{mathtools}

\usepackage[shortlabels]{enumitem}
\newcommand{\ECNU}{School of Physics and Technology, Nantong University, Nantong, 226019, People’s Republic of China}
\newcommand{\JSU}{Department of Physics, Jiangsu University, Zhenjiang 212013, China}
\begin{document}

\title{Enhancement in temperature sensing of a reservoir by Kerr-nonlinear resonator}
\author{Naeem Akhtar}
%\email{naeemakhtar166067@gmail.com}
\affiliation{\JSU}
\author{Jia-Xin Peng}
\email{18217696127@163.com}
\affiliation{\ECNU}
%\thanks{First and second authors contributed equally}
\author{Xiaosen Yang}
\email{yangxs@ujs.edu.cn}
\affiliation{\JSU}
\author{Yuanping Chen}
\email{chenyp@ujs.edu.cn}
\affiliation{\JSU}
\date{\today}
\begin{abstract}
The challenge of developing high-precision temperature sensors is an important issue that has recently received a lot of attention. In this work, we introduce an estimation technique to precisely measure the temperature of a quantum reservoir using a Kerr-nonlinear resonator with drive. Thermalization in our suggested protocol is assessed using Uhlmann-Jozsa fidelity, and then we utilize quantum Fisher information to evaluate the metrological potential of our thermometry scheme. We observe that increasing the Kerr nonlinearity coefficient and driving amplitude significantly enhances precision in the temperature estimation. Furthermore, we also explore the underlying physical mechanisms by analyzing probe purity in the steady state and evaluating the performance of homodyne versus heterodyne detection methods. Our results demonstrate that neither of these Gaussian measurements is optimal; instead, optimal homodyne detection always surpasses heterodyne detection.
\end{abstract}
\maketitle
%\section{Introduction} \label{I}

\paragraph*{Introduction.--}Quantum metrology~\cite{degen2017quantum,giovannetti2004quantum,gemmer2009quantum,giovannetti2011,giovannetti2006quantum}  has expanded the boundaries of thermodynamics~\cite{lewis2020thermodynamics,cahn1997thermodynamics,childs2000review,mcgee1988principles}, allowing small objects to be cooled to extremely low temperatures~\cite{bloch2008many,giazotto2006opportunities}.\;Quantum thermodynamics combines the thermodynamic principles and laws of quantum physics to explain the effects of thermalization, refrigeration, transport phenomena, out-of-equilibrium dynamics, and assessing the thermal properties of small-scale objects and thermal reservoirs~\cite{lewis2020thermodynamics,cahn1997thermodynamics,childs2000review,mcgee1988principles}. Measuring thermodynamic quantities at the quantum level necessitates tremendous precision~\cite{degen2017quantum} as well as superior cooling~\cite{timofeev2009electronic}. Temperature, defined through the zeroth law of thermodynamics, is one of the important thermodynamical parameters, and a precise measurement of temperature is crucial for both classical and quantum thermodynamics. Significant interest has been shown in the highly accurate temperature measurement of quantum reservoirs~\cite{PUGLISI20171,Hovhannisyan2021,Kirkova2021}. 

The precise measurement of temperature in a quantum system is challenging due to the limitations imposed by quantum principles~\cite{gemmer2009quantum,giovannetti2011,giovannetti2006quantum}. Quantum thermometry~\cite{PhysRevE.110.024132,10.1063/5.0207531, PRXQuantum.4.040314, PhysRevA.109.042417, PhysRevA.110.032605, haupt2014single,Dedyulin2022}, a promising field in quantum metrology, employs quantum thermometers to measure low temperatures with accuracy well beyond the standard bound set by classical statistics~\cite{giovannetti2011,giovannetti2006quantum}. Most thermometry schemes are based on a two-level system~\cite{cavina2018bridging,feyles2019dynamical} or a harmonic oscillator~\cite{hovhannisyan2018measuring,khan2021quantum}, working as the quantum sensor, is connected to the quantum reservoir of interest~\cite{zhang2021non}. The interaction between the sensor and reservoir encodes information about the temperature associated with the reservoir into the state of the sensor, which is then obtained by measuring a specific observable associated with the sensor.\;A number of techniques have been proposed to achieve a precise measurement of temperature associated with quantum systems, which included quantum control~\cite{mukherjee2019enhanced,kiilerich2018dynamical}, strong coupling~\cite{brenes2023multispin,correa2017enhancement}, non-Markovian effects~\cite{zhang2021non,xu2023non,zhang2022non}, and adding auxiliary devices~\cite{brunelli2012qubit,guo2015improved,ullah2023low}, to name a few.

In general, there are two distinct types of quantum thermometers that rely on different techniques. In the first instance, equilibrium thermometry~\cite{mehboudi2019thermometry,Stace2010,Campbell2018,Correa2015,Correa2017}, which is a technique for measuring temperature that is relatively more common, involves the contact of a probe with a sample (a reservoir of temperature $T$). The probe is then measured once it has reached equilibrium with the system and is thus defined by a thermal state at the same temperature as the system. Nonequilibrium thermometry is another thermometry scheme~\cite{brunelli2011qubit,brunelli2012qubit,Pasquale2017,Montenegro2020,feyles2019dynamical,gebbia2020two,Mancino2020,Mitchison2020,mukherjee2019enhanced,Seah2019}, in which temperature is extracted as a parameter from the state of the probe before its thermalization. In equilibrium thermometry, which is the focus of this work, energy measurement is found to be the optimal choice for saturating the quantum Cramér-Rao (CR) constraint~\cite{Stace2010} independent of temperature. In contrast, in nonequilibrium thermometry, the optimal measurement, which minimizes the uncertainty of the estimation by saturating the CR inequality~\cite{cramer1999mathematical}, is typically dependent on the unknown temperature of the system, making its realization difficult in practice.

In this work, we provide an estimation technique for precisely measuring the temperature of a thermal reservoir. Our estimation scheme is employed by using a nonlinear Kerr-resonator with drive to precisely estimate the temperature of a bosonic reservoir. Kerr-resonators have also been utilized in the past to improve the precision of quantum magnetometers and the accuracy of superconducting qubit readouts~\cite{di2023critical}. In this work, we anticipate that the Kerr-resonator thermometer enables high-precision temperature detection. Our study presents the physical mechanism by examining the probe's purity in the steady state, and then we also examine the performance of actual Gaussian measurements. We investigate the effects of the Kerr nonlinearity coefficient and external single-photon driving on sensing performance. Specifically, we estimate the metrological potential of the resulting Kerr resonator thermometer using theoretical verification such as quantum Fisher information, and then we observe that increasing the controlling parameters such as Kerr nonlinearity coefficient and driving amplitude significantly improves the temperature estimation precision of our method. To enhance our analysis, we also investigate the underlying physical mechanism by examining probe purity in the steady state, and we specifically evaluate the performance of homodyne and heterodyne detections, demonstrating that neither of the two standard Gaussian measurements is optimal; however, optimal homodyne detection is always superior to heterodyne detection.

%The structure of this paper is as follows: We present the Kerr thermometer model and review a few basic concepts in parameter estimation theory in Sec.~\ref{II}.  We examine the performance of a Kerr-resonator thermometer in Sec.~\ref{III}, mainly examining how driving and Kerr coefficient affect sensing accuracy. In Sec.~\ref{IV}, we examine whether actual Gaussian measurements can achieve the precision limit specified by QFI and here we compare the results of two considered typical Gaussian measurements. 
%The conclusion of this study is provided in Sec.~\ref{V}.

\begin{figure}[t!]
	\includegraphics[width=\columnwidth]{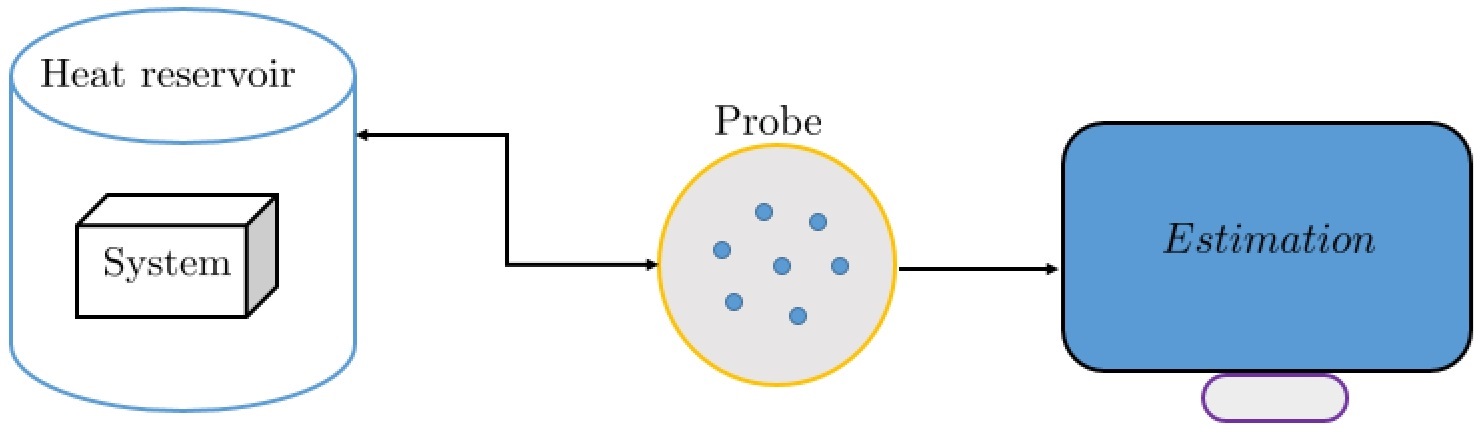} 
	\caption{{A general sensing approach for measuring temperature or any other parameter connected with a reservoir at thermal equilibrium.  This can be accomplished by temporarily connecting the sample with a probe; in the present case, the Kerr-nonlinear resonator is linearly coupled to a sequence of harmonic modes in a heat reservoir. These modes are in thermal equilibrium with a specific temperature ($T$). With this setting, the resonator serves as a probe for determining the temperature of the reservoir.}}
	\label{figModel}
\end{figure}
\begin{figure}[hbt!]
\centering
	\includegraphics[width=0.505\textwidth]{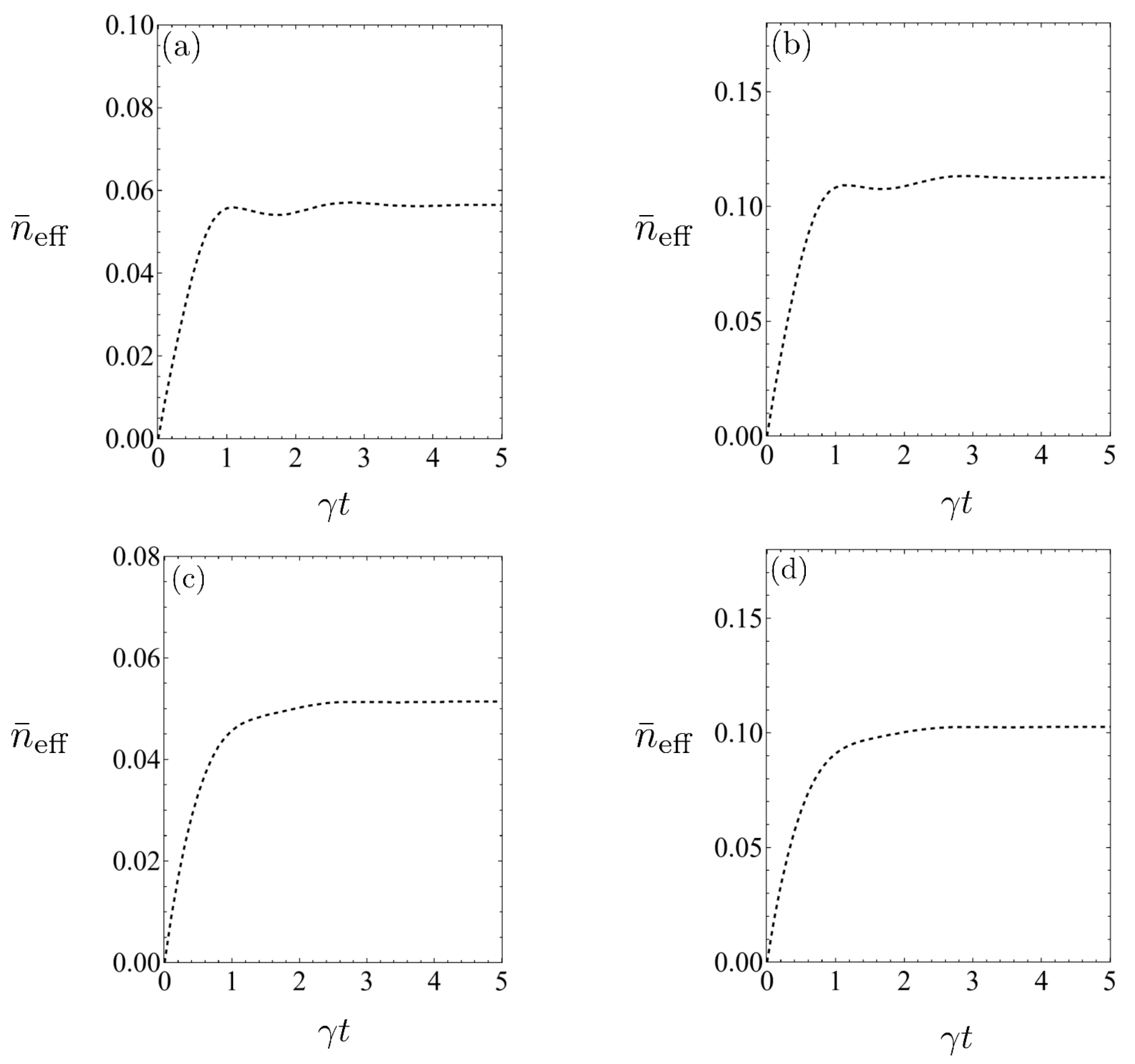} 
	\caption{$\bar{n}_{\text{eff}}$ represents the effective temperature reached by a Gibbs state in the illustration, as described in Eq.~(\ref{eq:neff}). (a)~$n_{\text{th}}=0.05$, $\chi=0.5\gamma$, $\delta=-3.5\gamma$, $\mathcal{E}=\gamma$, (b)~$n_{\text{th}}=0.1$, $\chi=0.5\gamma$, $\delta=-3.5\gamma$, $\mathcal{E}=\gamma$, (c)~$n_{\text{th}}=0.05$, $\chi=0.5\gamma$, $\delta=-3.5\gamma$, $\mathcal{E}=0.5\gamma$, and (d)~(c)~$n_{\text{th}}=0.1$, $\chi=0.5\gamma$, $\delta=-3.5\gamma$, $\mathcal{E}=0.5\gamma$.}
	\label{eff_tempo}
\end{figure}

\paragraph*{Scheme for Temperature Measurement.--}First, let us introduce the model of the system that is employed in the current work to precisely measure the temperature of the reservoir.
The generic model of a single-photon-driven dissipative Kerr-nonlinear resonator, which acts as our quantum probe for sensing environmental temperature.
The Hamiltonian of a single-photon-driven Kerr-nonlinear resonator of frequency $\omega_a$ in a rotating frame at the driving frequency $\omega_L$ is given by (we set $\hbar=1$ hereafter) \cite{asjad2023joint}
\begin{equation}\label{eq:model1}
\hat{H}_{\text{sys}}=\delta \hat{a}^\dagger \hat{a} +\chi \hat{a}^{\dagger}  \hat{a}^{\dagger}  \hat{a}\hat{a}  + \mathcal{E}(\hat{a}+ \hat{a}^\dagger), 
\end{equation}
where $\delta=\omega_a-\omega_L$ is the cavity-pump detuning, $\chi$ denotes the Kerr-nonlinearity strength, and $\mathcal{E}$ is the amplitude of the single-photon drive. Here, $\hat{a}$ and $\hat{a}^\dagger$ are the annihilation and creation operators for photons inside the resonator, respectively. It is worth mentioning that the experimental implementation of the Kerr-nonlinear resonator has been widely adopted. For example, a typical scheme is to couple the resonator to a superconducting quantum interference device (SQUID), and then the SQUID can induce Kerr nonlinearity~\cite{lin2014josephson,krantz2013investigation}. Figure~\ref{figModel} depicts a schematic representation of such a setup, which can be used to execute our suggested Kerr-nonlinear resonator-based thermometery system.

\begin{figure*}[hbt!]
	\includegraphics[width=1\textwidth]{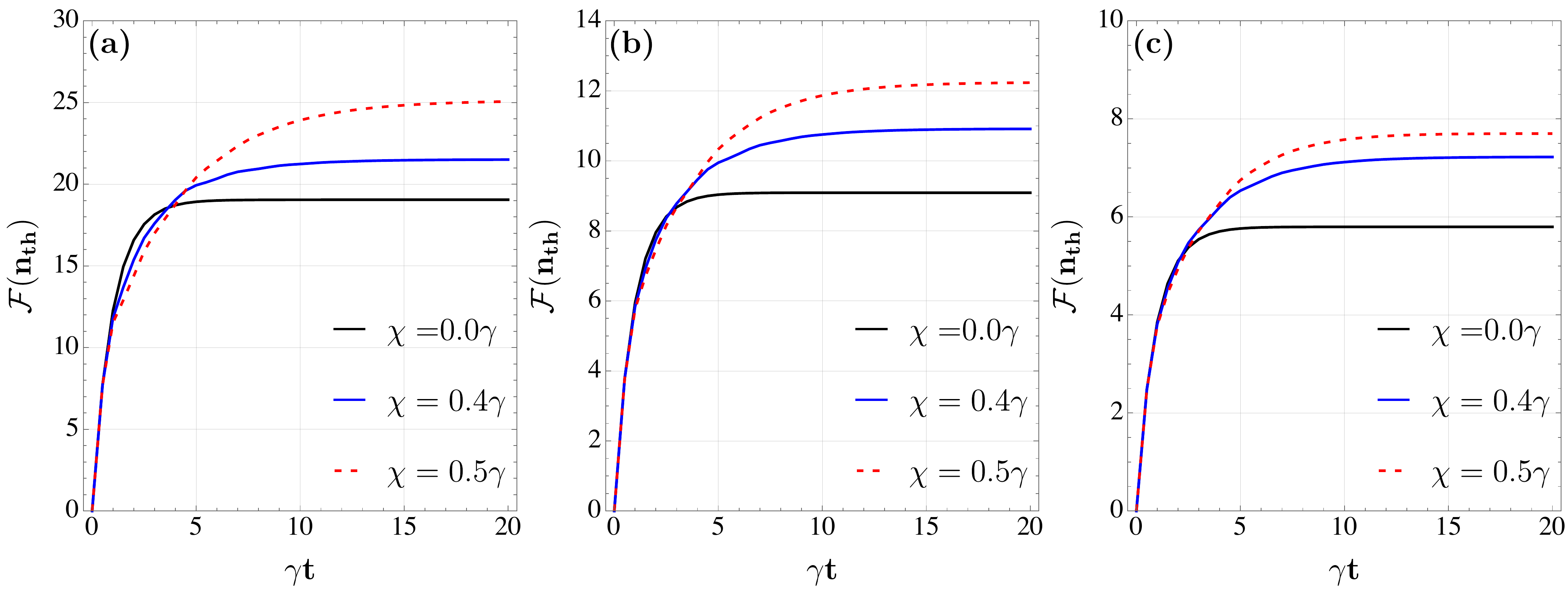} 
	\caption{The dynamic evolution  of $\mathcal{F}$($n_{\text{th}}$) with different Kerr-nonlinearity coefficients, where (a) $n_{\text{th}}=0.05$; (b) $n_{\text{th}}=0.1$; (c) $n_{\text{th}}=0.15$. The detuning $\delta=-3.5 \gamma$ and drive amplitude $\mathcal{E}=\gamma$. }
	\label{Fig1}
\end{figure*}

The Kerr-nonlinear resonator is linearly coupled to series of harmonic modes of a heat bath. These modes are in a multimode thermal equilibrium state with a well-defined temperature $T$. With this setting, the resonator act as probe for assessing the temperature of the bath. The total Hamiltonian of the probe and the bath is given by
\begin{equation}\label{eq:model}
\hat{H}_{\text{tot}}=\hat{H}_{\text{sys}}+ \sum\limits_{k=1} [\omega_{k} \hat{\nu}^\dagger_k \hat{\nu}_k + g_{k} (\hat{\nu}^\dagger_k \hat{a} + \hat{a}^\dagger \hat{\nu}_k)],
\end{equation}
where $g_k$ quantifies the probe-bath coupling strength and $\hat{\nu}_k$ ($\hat{\nu}^\dagger_k$) is the annihilation (creation) operator of the $k$th harmonic oscillator of the bath, with frequency $\omega_k$.

Suppose the whole initial joint state of the probe and bath is $\hat{\rho}(0)\otimes\hat{\rho}_B$, where $\hat{\rho}(0)$ is a density operator of the Kerr resonator and $\hat{\rho}_B$ is the canonical Gibbs state of the thermal bath~\cite{mehboudi2019thermometry}. Since we assess the information  of the bath temperature as imprinted on the quantum probe, we only focus on the dynamics of the Kerr-nonlinear resonator. When the coupling strength $g_k$ is weak, by tracing over the degrees of freedom in the heat bath, we ultimately get reduced master equation for Kerr-nonlinear resonator~\cite{isar1994open,breuer2002theory}, that is,
\begin{equation}
\label{Eqq}
\dot{\hat{\rho}}(t)= -i[\hat{H}_{\text{sys}},\hat{\rho}(t)] + \gamma (n_{\text{th}} +1) \mathcal{\hat{D}}(\hat{a}) \hat{\rho}(t) 
+ \gamma n_{\text{th}} \mathcal{\hat{D}}(\hat{a}^\dagger) \hat{\rho}(t).  
\end{equation}
To solve this equation, we use Markovian-approximation and Gibbs distribution for bath oscillators; $\mathcal{\hat{D}}(\mathcal{J})\hat{\rho}= 2 \mathcal{\hat{J}}\hat{\rho} \mathcal{\hat{J}}^\dagger-\mathcal{\hat{J}}^\dagger \mathcal{\hat{J}}\hat{\rho}-\hat{\rho}\mathcal{\hat{J}}^\dagger\mathcal{\hat{J}}$ is the Linblad dissipation superoperator, $\mathcal{\hat{J}}$ is the quantum jump operator, $\gamma$ is the corresponding decay rate, and $n_{\text{th}} =( e^{\nicefrac{\omega_a}{k_B T}}-1)^{-1}$ denotes the mean thermal photon number associated with the resonator at thermodynamic equilibrium temperature $T$ and $k_{B}$ is the Boltzmann constant. As a result of the interaction between the environment and the Kerr resonator, the reduced density matrix of the Kerr-nonlinear resonator contains information on the corresponding temperature $T$.

A thermal state, referred to as the Gibbs state, is denoted as
\begin{align}
  \hat{\rho}_{Gibbs} =\frac{1}{(\bar{n}_{\text{eff}}+1)} \sum_{j=0}^{+\infty} \Big(\frac{\bar{n}_{\text{eff}}}{\bar{n}_{\text{eff}}+1}\Big)^j   \ket{j}\bra{j}  .\end{align} 
  The closeness between the Gibbs state $\hat{\rho}_{Gibbs}$ with the corresponding state of our system, denoted as $\hat{\rho}(t)$ can be evaluated by using the quantity known as Uhlmann-Josza fidelity (UJF)~\cite{liang_quantum_2019}. Generally, this quantity for two density matrices $\hat{\rho}$ and $ \hat{\sigma}$ can be analyzed by using,
\begin{equation}
O\left(\hat{\rho},\hat{\sigma}\right) = \left(\mathrm{Tr} \sqrt{\sqrt{\hat{\sigma}}\hat{\rho} \sqrt{\hat{\sigma}}}\right)^2  = 	O \left(\hat{\sigma},\hat{\rho}\right),
\label{eq:fidelity}
\end{equation}
and it holds the following properties: First, $O= 1$ when $\hat{\rho} = \hat{\sigma}$; second, if $\hat{\rho}$ and $\hat{\sigma}$ are pure states, the formula simplifies to $O = \mathrm{Tr}(\hat{\rho}\hat{\sigma})$, which is just a simple overlap between two states; thirdly, $0\leq O \leq 1$; and it is invariant under any unitary transformation.

We numerically evaluate the UJF for the density matrix $\hat{\rho}(t)$ and the corresponding Gibbs state $\hat{\rho}_{Gibbs}$. In our approach, the UJF protocol ensures that the system undergoes near thermalization. This is confirmed by extracting the effective temperature $\bar{n}_{\text{eff}}$ from
\begin{equation}
\bar{n}_{\text{eff}}\sim\arg_{\max_{\bar{n}_{\text{eff}}}} O[\hat{\rho}(t), \hat{\rho}_{\text{Gibbs}}(\bar{n}_{\text{eff}})], 
\label{eq:neff}
\end{equation}
where $\bar{n}_{\text{eff}}$ is a parameter associated with the temperature of the Gibbs state. The maximization of the fidelity function $O$ identifies the value of $\bar{n}_{\text{eff}}$ that results in the best match between the evolved state $\hat{\rho}(t)$ and the Gibbs state $\hat{\rho}_{\text{Gibbs}}(\bar{n}_{\text{eff}})$.

This procedure finds the temperature $\bar{n}_{\text{eff}}$ that aligns most closely with the temperature of the subsystem $\hat{\rho}(t)$, which should approach the temperature of the reservoir $n_{\text{th}}$. In other words, by tracking the evolution of the temperature-related parameter of the subsytem, $\bar{n}_{\text{eff}}$ is optimized to keep the highest value of the fidelity, ensuring a close match between the state of the subsystem and the Gibbs state.

We verify the thermalization process by plotting the effective temperature $\bar{n}_{\text{eff}}$ as a function of time, as shown in Fig.~\ref{eff_tempo}. For each case, the corresponding curve $\bar{n}_{\text{eff}}$ indicates that the temperature of the subsystem asymptotically approaches the reservoir temperature $n_{\text{th}}$, confirming that the system has thermalized. The curves indicate that the effective temperature of the subsystem converges to  $n_{\text{th}}$, consistent with the expected outcome of thermalization.

\begin{figure}[hbt!]
	\includegraphics[width=0.3\textwidth]{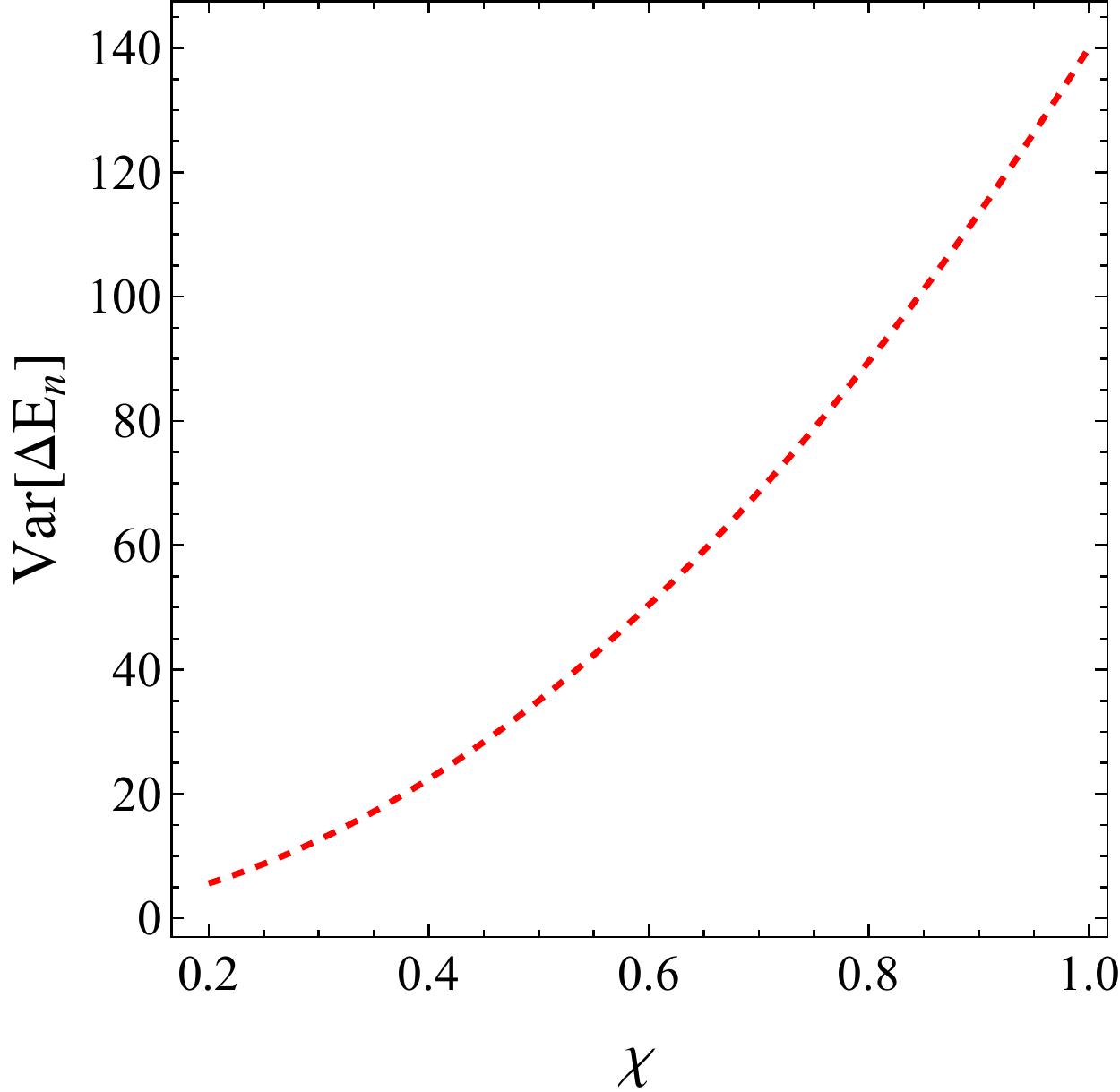} 
	\caption{The energy variance of the spectral gap of the system over the Kerr-nonlinearity coefficient with the energy levels under consideration varies from $n=30$ to $n=50$. The system parameters are $\delta=-3.5 \gamma$ and $\mathcal{E}=1$.}
	\label{fig:vari1}
\end{figure}

\begin{figure*}[hbt!]
	\includegraphics[width=1\textwidth]{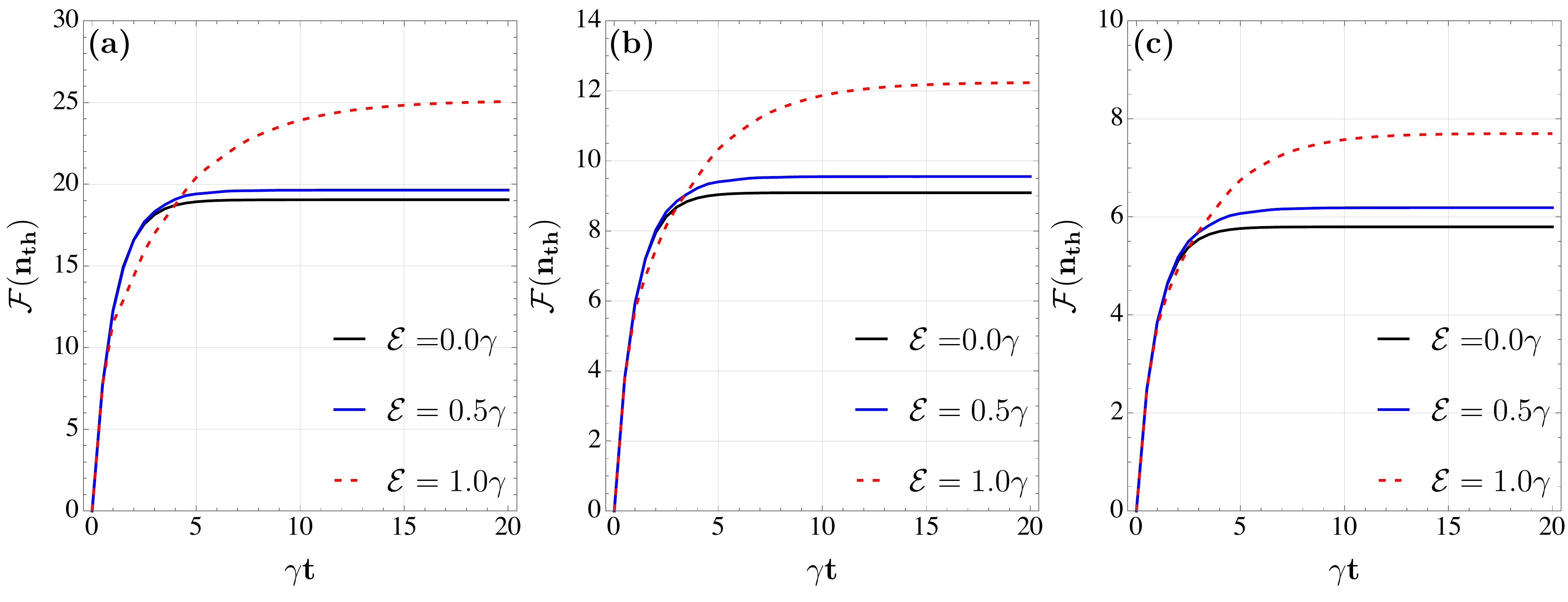} 
	\caption{The dynamic evolution of $\mathcal{F}$($n_{\text{th}}$) with different drive amplitudes, where (a) $n_{\text{th}}=0.05$; (b) $n_{\text{th}}=0.1$; (c) $n_{\text{th}}=0.15$.  The detuning $\delta=-3.5 \gamma$ and  Kerr nonlinearity coefficient $\chi=0.5 \gamma$.}
	\label{Fig2}
\end{figure*}

\paragraph*{Parameter Estimation Theory.--}The goal of quantum parameter estimation is to identify one or more specific values by conducting suitable measurements and utilizing an estimator algorithm~\cite{giovannetti2004quantum,gemmer2009quantum}. Here, we concentrate on the estimation of a single parameter, specifically the temperature $T$ of the thermal reservoir. It is assumed that all other parameters are controlled and already known. Let us consider a situation where we achieve the actual value $T$ of a thermal reservoir by building an unbiased estimator. For convenience, we estimate the precision in the average thermal photon number ($n_{\text{th}}$) using our proposed scheme, and then the corresponding information on the temperature ($T$) can be simply acquired because the $T$ is proportional to $n_{\text{th}}$, and this stay true for our cases under consideration. Hence, the corresponding estimation precision can be quantified through the mean-squared error of estimator denoted by $\Delta n_{\text{th}}^2$, which satisfies CR inequality \cite{liu2020quantum,fisher1925theory,helstrom1969quantum,holevo2011probabilistic,paris2009quantum,sha2022continuous}
\begin{equation}
\label{Eq4}
\Delta n_{\text{th}}^2\geq \frac{1}{\mu F(n_{\text{th}})},
\end{equation}
where $\mu$ is the number of independently repeated measurements and $F(n_{\text{th}})$ represents the classical Fisher information (CFI)~\cite{liu2020quantum,fisher1925theory,helstrom1969quantum,holevo2011probabilistic,paris2009quantum,sha2022continuous}, which is defined as
\begin{equation}
\label{Eq5}
F(n_{\text{th}})=\int  P(x|n_{\text{th}})   \left[ \partial_{n_{\text{th}}} \ln \{P(x|n_{\text{th}}) \}\right]^2dx,
\end{equation}
where $\partial_{n_{\text{th}}}:=\nicefrac{\partial}{\partial n_{\text{th}}}$, and $P(x|n_{\text{th}}) $ represents the conditional probability density of a measurement outcome $x$. When the estimator is optimal, the Eq.~(\ref{Eq4}) achieves equality. In the asymptotic regime, where the data set is very large, it has been demonstrated that the Bayesian or maximum likelihood  algorithm produces the most accurate estimator. The classical CR inequality mentioned above can be made more generalized by optimizing over all possible Positive-Operator-Valued-Measure (POVM) $\{\hat{\Pi}_x \}$  operators with $\sum \limits_{x} \hat{\Pi}_x\hat{\Pi}^\dagger_x = {\Bbb I}$ \cite{helstrom1969quantum,holevo2011probabilistic,paris2009quantum,sha2022continuous}. Consequently, we can get a new and improved CR lower bound, that is
\begin{equation}
\Delta n_{\text{th}}^2\geq \frac{1}{ \mu \mathcal{F}(n_{\text{th}})},
\end{equation}
which is known as quantum CR inequality. Here $\mathcal{F}(n_{\text{th}})$ is the QFI and can be written as 
\begin{equation}
\mathcal{F}(n_{\text{th}}):= \max_{\hat{\Pi}_x}F(n_{\text{th}})= \mathrm{Tr}[\hat{\rho}(n_{\text{th}}) \hat{L}^2],
\end{equation} 
where $\hat{L}$ is the symplectic logarithmic derivative (SLD) operator defined such that  $2\partial_{n_{\text{th}}} \hat{\rho}
(n_{\text{th}}) =\{ \hat{L},\hat{\rho} (n_{\text{th}}) \}$ \cite{helstrom1969quantum,holevo2011probabilistic,paris2009quantum,sha2022continuous}.

\begin{figure}
	\includegraphics[width=0.35\textwidth]{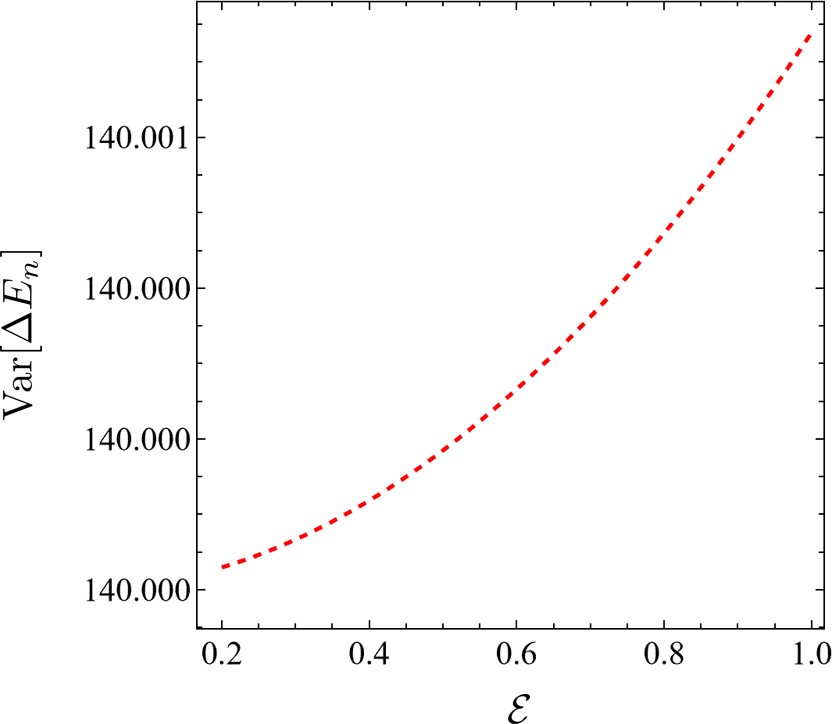} 
	\caption{The energy variance of the spectral gap of the system over the drive parameter with the energy levels under consideration varies from $n=30$ to $n=50$. The system parameters are $\delta=-3.5 \gamma$ and $\chi=1$.}
	\label{vari2}
\end{figure}

The sensing precision indicates the sensitivity of the system to estimated parameters, so that the calculation of Fisher information involves derivative operations for the unknown parameters. Furthermore, it is very difficult to analytically achieve the solutions of Eq.~(\ref{Eqq}). Therefore, we adopt numerical methods to solve this master equation, and then we study the sensing precision of the temperature of the bath. Note that we use Python built-in packages QuTiP to achieve our numerical results~\cite{johansson2012qutip}. As a technical remark, performing the numerical calculations, the first-order derivative of any function $f_{n_{\text{th}}}$ with respect to $n_{\text{th}}$ is treated by adopting the finite-difference method~\cite{sha2022continuous}, that is,
\begin{align}
&\nonumber\frac{\partial f_{n_{\text{th}}}}{\partial n_{\text{th}}}:\approx \\& \frac{%
	-f_{n_{\text{th}}+2dn_{\text{th}}}+8f_{n_{\text{th}}+dn_{\text{th}}}-8f_{n_{\text{th}}-dn_{\text{th}}}+f_{n_{\text{th}}-2dn_{\text{th}}}}{12dn_{\text{th}}},
\end{align}%
where, we set $\nicefrac{dn_{\text{th}}}{n_{\text{th}}}=10^{-7}$, which provides high numerical accuracy.

\begin{figure}[hbt!]
	\includegraphics[width=1\columnwidth]{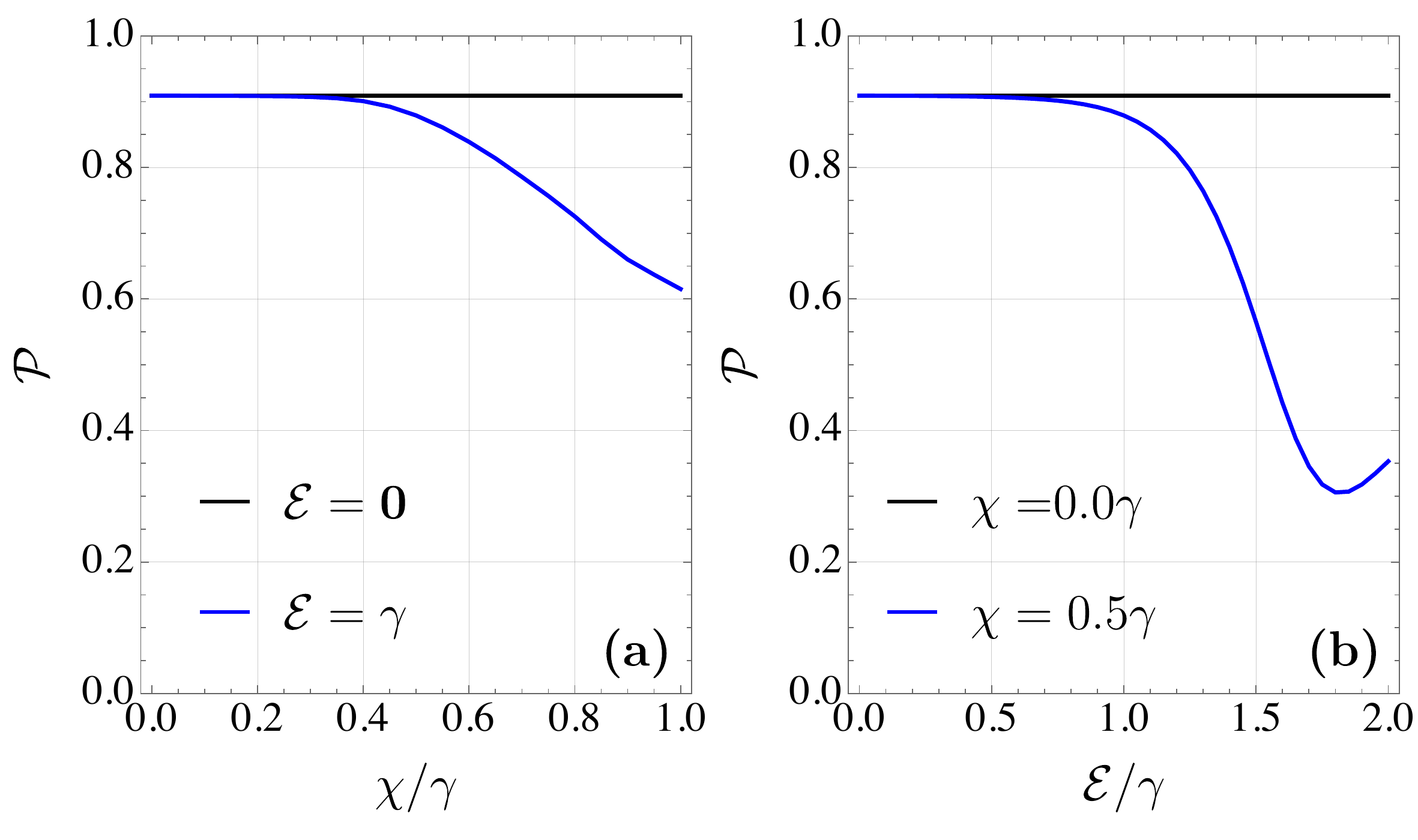} 
	\caption{The purity of steady state as a function of (a) Kerr-nonlinearity coefficient and (b) drive amplitude and, where 
		$n_{\text{th}}=0.05$ and the detuning $\delta=-3.5$.}
	\label{Fig3}
\end{figure}

\begin{figure*}[hbt!]
	\includegraphics[width=2\columnwidth]{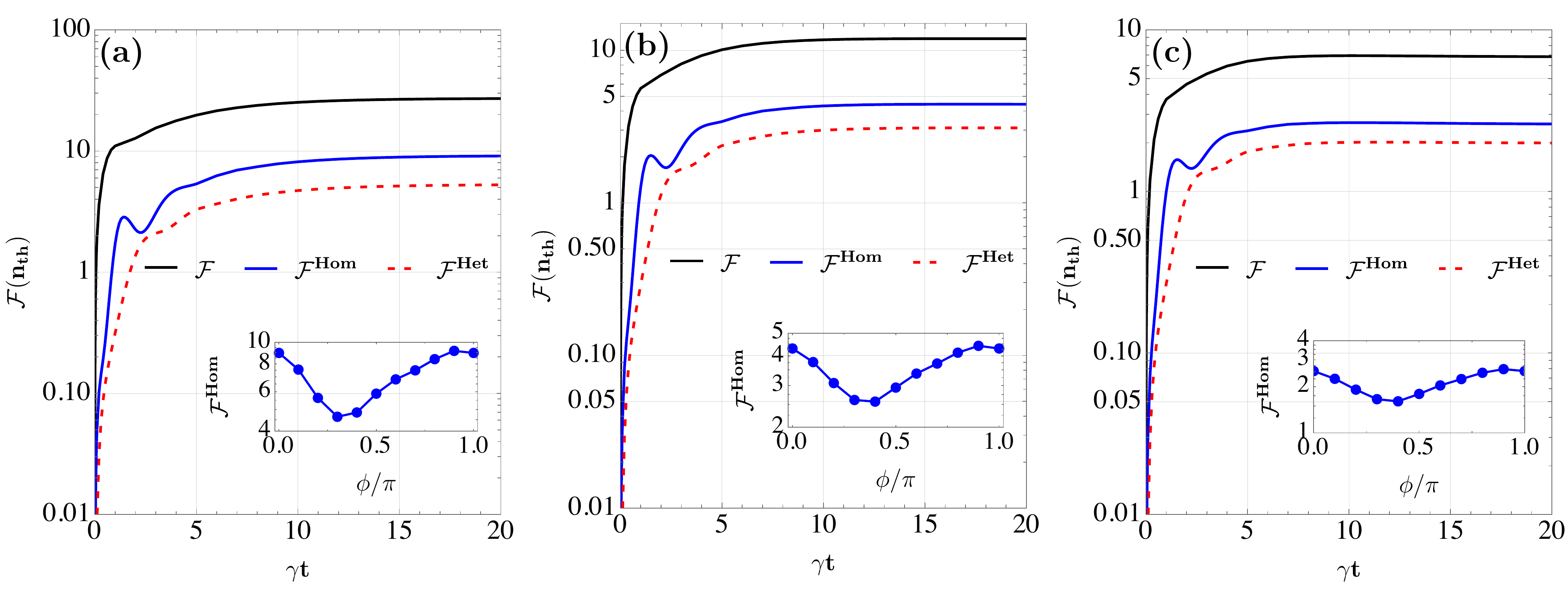} 
 \caption{Performance analysis of Gaussian measurement. $\mathcal{F}$($n_{\text{th}}$) (black), $\mathcal{F}^{\rm Hom}$ (blue) and  $\mathcal{F}^{\rm Het}$ (red)  as a function of $\gamma t$, where  (a) $n_{\rm th} =0.05$, (b) $n_{\rm th} =0.1$ and (c) $n_{\rm th} =0.15$ with $\delta=-3.5 \gamma$, $\chi=0.65 \gamma$, $\phi=0.9\pi$, and $\mathcal{E}=1.0\gamma$.}
	\label{Fig5}
\end{figure*}

\paragraph*{Enhancement in Temperature Sensing.--}In general, the precision in the estimation is quantified by the available Fisher information associated with the parameter that we aim to estimate. Here we explore the performance of the thermometer by quantifying the corresponding QFI $\mathcal{F}$($n_{\text{th}})$~\cite{PhysRevLett.132.060801,Aybar2022criticalquantum}. In particular, we focus on how the driving and Kerr nonlinearity strengths affect the temperature estimation. Throughout this analysis, we consider that the probe is initially in a vacuum state. 

Let us now explain our results. First, in Fig.~\ref{Fig1}, we provide QFI as a function of probe-bath interaction time for different values of the nonlinearity strengths $\chi$ and the temperature-associated parameters $n_{\text{th}}$. We utilize a rescaled version of the probe-bath interaction time ($\gamma t$) throughout the work. Here, we show that raising the Kerr-nonlinearity coefficient leads to more precise temperature estimation, as evidenced by a rapid increase in the QFI over $\gamma t$ in Figs.~\ref{Fig1} (a)-(c). This enhancement in QFI is clearly visible by raising $\chi$ for different values of $n_{\text{th}}$. Hence, these results reflect that introducing and enhancing the Kerr-nonlinearity coefficient allows a better estimation of the temperature. This is because, compared to linear oscillators, the energy level spacing of Kerr nonlinear oscillators varies nonlinearly. The impact of nonlinearity contained in the spectral gap of a system appeared to have an impact on the thermal QFI, as illustrated in~\cite{Correa2015,mok2021optimal,Campbell2018}.  The difference between the nearest neighbor energy levels $\Delta E_n $ is
\begin{equation}
\Delta E_n:  = E_{n+1} - E_{n},
\end{equation}
where $E_n$ is reflecting energy spectra associated with our system. The variance of a collection of $\Delta E_n$ can written as
 \begin{equation}
 \mathrm{Var}[\Delta E_n] := \frac{1}{N-1} \sum_{i=1}^N\left(\Delta E_i- \overline{\Delta E}\right)^2,
 \end{equation}
 where $\overline{\Delta E}$ is the average value $\overline{\Delta E} = \nicefrac{1}{N} \sum_{i=1}^N \Delta E_i$
We observe the spectral variance of the spectral gap, denoted as $\text{Var}[\Delta E_n]$. The corresponding $\text{Var}[\Delta E_n]$ is plotted in Fig.~\ref{fig:vari1} over the parameter $\chi$, where the variances for a certain higher energy levels clearly show nonlinearity present in the spectral gap over the given parameter. Therefore, for disturbances that are insensitive to linear oscillators, Kerr oscillators may be very sensitive. This is why the Kerr oscillator can improve estimation precision. In our method, the interaction with the environment is the primary source of encoding temperature information, but it can also impair exact measurement due to decoherence caused by this interaction~\cite{isar1994open,breuer2002theory,akhtar2024}, implying that there is a trade-off between these two effects. However, when QFI reaches its flat levels, both decoherence effects and the encoding process approach equilibrium, and so sharp variations in QFI are minimized. Moreover, in the long time limit, the resonator is completely thermalised with the bath due to the incoherent Linblad dynamics~\cite{CHANDA20161}, and hence QFI saturates to a fixed value once the probe is fully thermalised. For each of the cases presented in Figs.~\ref{Fig1} (a)-(c), it is interesting to note that the saturation of the corresponding QFI to attain a constant amplitude is dependent on the values of the parameter $\chi$; that is $\chi\neq 0$, the probe takes a comparatively longer time to fully thermalise.

Figures~\ref{Fig1} (a)-(c) show that larger values of the parameter $\chi$ result in slower curves approaching flat levels, indicating higher QFI (or more accurate temperature measurement). 
In our cases, the non-linear Kerr-resonator thermometer ($\chi\neq 0$) has a sufficiently higher sensing precision than that of the linear resonator thermometer ($\chi=0$). Different thermometric techniques for relevant experimental setups have been realized to be dependent on ultimate bounds and scaling laws, which limit the precision of temperature estimates for systems in and out of thermal equilibrium~\cite{mehboudi2019thermometry}. For example, in some cases, obtaining temperature information in a short period of time is a significant challenge in experiments, and it has been discovered that achieving high-precision temperature sensing over a long period of time is advantageous for such temperature estimation schemes based on kerr-nonlinear resonator types.
 In addition, by comparing the corresponding cases depicted in Figs.~\ref{Fig1}(a)-(c), we also observe that for same values of $\chi$, increasing temperature does not necessarily result in greater estimation precision. These effects may be aroused due to the thermal noise associated with the temperature of the reservoir~\cite{forsythe1999dissipative}, where the temperature of the bosonic reservoir may have negative impact on coherence~\cite{isar1994open,breuer2002theory}, that is, higher temperatures leading to higher decoherence, which effects on the estimation precision.

We now present $\mathcal{F}$($n_{\text{th}}$) for a few driving amplitudes $\mathcal{E}$ and is plotted against $\gamma t$ as shown in Fig.~\ref{Fig2}. As observed in Fig.~\ref{Fig2}, increasing the driving amplitude $\mathcal{E}$ results in higher QFI, indicating that increasing the driving amplitude intensity $\mathcal{E}$ enhances temperature sensing precision. For this situation, corresponding $\text{Var}[\Delta E_n]$ is plotted over $\mathcal{E}$ in Fig.~\ref{vari2} capturing the nonlinearity present in the spectral gaps associated with the higher energy levels. This is another physical indication of how the parameter $\mathcal{E}$ provides a positive impact on the QFI of our system. Moreover, physically, the increase in driving can significantly increase the average photon number of the probe  [see Eq.~(\ref{eq:model1})], thereby increasing the number of carriers for storing temperature information and achieving an improvement in estimation precision. This is because a higher photon number in the system may lead to enhanced metrological capacity of a system~\cite{Lee_2019}.

To provide a more physical description of the improvement in temperature estimation due to the Kerr coefficient and driving amplitude, we depict the steady state purity: $ \mathcal{P}=\text{Tr}(\left[\hat{\rho}\left(t\rightarrow \infty \right) \right] ^{2}$ as a function of $\chi$ and $\mathcal{E}$ in Figs.~\ref{Fig3} (a) and (b), respectively. Note that from a geometric perspective, QFI also represents the distinguishability of adjacent quantum states of different purity values, which implies that QFI is closely related to fidelity~\cite{yu2022}. Figure~\ref{Fig3} shows that when $\mathcal{E}$ ($\chi$) is zero, then purity of the steady-state is constant over $\chi$($\mathcal{E}$). On the other hand, as depicted in Figs.~\ref{Fig3} (a) and (b), raising the driving amplitude and Kerr coefficient both result in a decrease in the purity of probe steady state, i.e., thermalizing degree of the Kerr-resonator thermometer is more significant. This provides physical reasons for how our two processes improve temperature sensing.

\paragraph*{Homodyne versus Hetrodyne Measurements.--}In general, performing optimal measurements, that allow for achieve optimal bound for QFI, are hard to implement. In contrast, Gaussian measurements such as homodyne and hetrodyne are generally less demanding and these can be easily implemented in the experiments~\cite{yuen1983noise,scully1999quantum,agarwal2012quantum}. Such measurement are readily realizable for optical mode~\cite{Collett1987}. Additionally, methods have been developed to perform Gaussian measurements for microwave modes \cite{Meinel2021}. In this section, we analyze the performance of homodyne and heterodyne detection scheme for the present thermometery scheme. Specifically, we evaluate the CFI of Gaussian measurement scheme for ambient temperature and compare its value with the corresponding QFI. We now discuss these measurements protocol and the corresponding positive operator value measurement (POVM).

In the homodyne detection scheme, it is required to measure the quadrature $\hat{Q}_{\theta }$ of the light field as given by~\cite{aspelmeyer2014cavity}
\begin{equation}
\hat{Q}_{\phi }=\frac{1}{2}\left( \hat{a}e^{-i\phi }+\hat{a}^{\dag
}e^{i\phi }\right).
\end{equation}%
The corresponding POVM is composed of the elements $\{ |\hat{Q}%
_{\phi }^{i}\rangle\} $ of $\hat{Q}_{\phi }$, that is,

\begin{align}
   \{ \hat{%
	\Pi}_{\hat{Q}_{\phi }}^{i}=|\hat{Q}_{\phi }^{i}\rangle \langle \hat{Q}%
_{\phi }^{i}|\},\;\text{with}\;\hat{Q}_{\phi }|\hat{Q}_{\phi
}^{i}\rangle =Q_{\phi }^{i}|\hat{Q}_{\phi }^{i}\rangle.
\end{align}
This implies that when the measurement result is $Q_{\phi }^{i}$, the corresponding
probability distribution is $P(Q_{\phi }^{i}|n_{\text{th}}) =$
Tr$[ \hat{\rho}(t) Q_{\phi }^{i}] $. By combining
Eq.~(\ref{Eq5}) with $P(Q_{\phi }^{i}|n_{\text{th}}) $, the CFI for
homodyne detection can be obtained. 
For heterodyne detection, the measurement projection operators constructed
from the coherent state is $\{ \hat{\Pi}^{\alpha }=\nicefrac{|\alpha \rangle
\langle \alpha |}{\pi}\}$. Similarly, the probability distribution of
the measurement result under heterodyne detection can be obtained as $%
P\left( \alpha |n_{\text{th}}\right) =$ Tr$[ \hat{\rho}( t) 
\hat{\Pi}^{\alpha }]$. Furthermore, the  formula (\ref{Eq5}) can be used to obtain CFI under heterodyne detection.

Finally, we compare the precision limit provided by QFI to Gaussian detection approaches.\;As illustrated in Fig.~\ref{Fig5}, we observe that the CFI obtained through Gaussian measurement is smaller than the QFI. This means that neither of the two typical Gaussian measurements is an optimal measurement strategy. Interestingly, we can observe that the optimal homodyne detection is always superior to heterodyne detection. The embedded  small figures show the CFI under different homodyne detections to determine the optimal homodyne detection. Particularly, homodyne detection for
amplitude quadrature ($\phi=0$) is always surpasses the homodyne detection for phase
quadrature ($\phi=\nicefrac{\pi}{2}$).

In summary, we have presented a temperature estimation scheme, in which the temperature of the quantum reservoir is measured using a Kerr-nonlinear resonator as a probe. We analyzed the impact of the Kerr nonlinearity coefficient as well as external single-photon driving on the sensing performance, and found that the estimation precision can be significantly improved by increasing nonlinear strength and driving amplitude. Furthermore, we evaluated the actual Gaussian measurement performance. The results indicate that the precision limit provided by QFI cannot be achieved based on two typical Gaussian measurements. Compared to heterodyne detection, the optimal homodyne is superior.  We believe that our investigation may helpful for the understanding and design of reservoir engineering.

As a final remark, we found that both the Kerr nonlinearity parameter and the driving amplitude positively influence the energy spectral gap of our system. Specifically, higher values of these parameters correspond to higher energy levels, which, in turn, lead to better optimization of the quantum Fisher information (QFI). However, it is important to note that achieving the optimal QFI at thermalization is a challenging task. In particular, optimizing the probe in its excited state remains difficult. The development of optimized probes for achieving the best QFI at thermal equilibrium is a critical area of the research~\cite{Correa2015}, which can be further explored for our present work in the future. Addressing this challenge may have significant implications for the creation of highly efficient quantum probes, which will be valuable for practical applications in quantum technologies.

\paragraph*{Acknowledgment.--} The authors express their sincere gratitude to Muhammad Asjad (at Simon Fraser University), H. K. Lau (at Simon Fraser University), and Muhammad Miskeen Khan (at University of Colordo) for insightful comments and suggestions on our work, which helped us to improve its quality.
\bibliography{Refn}
\end{document}